\documentclass[12pt]{iopart}
\newcommand{\BEQ}{\begin{equation}}     
\newcommand{\BEA}{\begin{eqnarray}}
\newcommand{\EEQ}{\end{equation}}       
\newcommand{\EEA}{\end{eqnarray}}

\begin{document}

\input epsf.sty

\title{Characterisation of non-equilibrium growth through global two-time quantities}

\author{Yen-Liang Chou$^1$ and Michel Pleimling$^1$}
\address{$^1$Department of Physics, Virginia Polytechnic Institute and State University, Blacksburg, Virginia
24061-0435, USA}
\eads{\mailto{ylchou@vt.edu}, \mailto{Michel.Pleimling@vt.edu}}

\begin{abstract}
In order to characterise non-equilibrium growth processes, we study the behaviour of global quantities that
depend in a non-trivial way on two different times. We discuss the dynamical scaling forms of global
correlation and response functions
and show that the scaling behaviour of the global response can depend on
how the system is perturbed. On the one hand we derive exact expressions for systems characterised
by linear Langevin equations (as for example the Edwards-Wilkinson and the noisy Mullins-Herring equations),
on the other hand we discuss the influence of non-linearities on the
scaling behaviour of global quantities by integrating numerically the Kardar-Parisi-Zhang
equation. We also discuss global fluctuation-dissipation ratios and how to use them for the characterisation of
non-equilibrium growth processes.
\end{abstract}
\pacs{05.70.Np,81.15.Aa,64.60.Ht}
\maketitle

\section{Introduction}
Non-equilibrium growth is ubiquitous in nature and is encountered in fields as diverse as materials science
or biological physics \cite{Mea93,Hal95,Bar95}. Kinetic roughening of interfaces displays a high degree 
of universality which has been the focus of many theoretical studies \cite{Kru97,Kru95}. 

Systems with non-equilibrium growth typically have three different dynamic regimes. Starting from a flat substrate,
the surface initially grows in an uncorrelated way, yielding what is sometimes called the random deposition (RD) regime.
As time goes on, correlations build up, leading to the correlated regime. Even though the nature of this regime depends on the
rules of the growth process, one generically finds that the time dependent mean interface width $W(t)$
increases as a power of time, $W(t) \sim t^\beta$,
where $\beta$ is the so-called growth exponent. This correlated growth continues until the saturation regime is reached
for which $W \sim L^\alpha$ where $L$ is the linear size of the system and $\alpha$ is the roughness exponent.
The different growth universality classes are characterised by different values of the exponents $\alpha$ and $\beta$.
In the past most studies focused on the scaling behaviour of the
surface width, as exemplified by the celebrated Family-Vicsek scaling relation \cite{1985FF,Chou09a}.
It is only rather recently (however, see \cite{Kal99} for an early exception) that the study of correlation
and response functions has been
shown to yield interesting insights into non-equilibrium growth processes and the related
interface fluctuations \cite{Rot06,Rot07,Bus07a,Bus07b,Chou09b,Ngu09,Noh10}.

Two-time quantities are nowadays routinely studied in the context of non-equilibrium relaxation and ageing
phenomena \cite{HenPle} where in most cases the focus is on local quantities. For example when studying a magnetic
system one usually investigates the spin-spin correlation function or the response of a spin to a local magnetic field.
Incidentally, the study of ageing in magnetic systems has revealed that additional insights can be gained
by looking at global quantities, as for example the magnetisation-magnetisation correlation or the response
of the magnetisation to a spatially constant magnetic field \cite{May03,Ple05,Ann06,Cal06,Dut09}.

In the studies of growth processes mainly local two-time quantities have been investigated in the past
\cite{Kal99,Rot06,Rot07,Bus07a,Bus07b,Chou09b,Ngu09,Noh10}. Examples include the two-time height-height correlation
function \cite{Kal99,Rot06,Rot07,Bus07a,Bus07b,Noh10} or the response of the height to a local perturbation 
\cite{Rot06,Rot07,Bus07a,Bus07b,Noh10}. Some attention has also been paid to slightly more complex quantities as
for example the two-time roughness or the two-time incoherent scattering function \cite{Bus07a,Bus07b}. 

Both the local height-height autocorrelation function $C_\ell(t,s)$ and the local autoresponse function $R_\ell(t,s)$ display simple ageing scaling
forms \cite{Kal99,Rot06,Rot07,Noh10}, irrespective of whether the system is linear or non-linear:
\begin{equation}
C_\ell(t,s) = s^{-b_\ell} \, f_{C_\ell}(t/s) ~~~~, ~~~~ R_\ell(t,s) = s^{-1-a_\ell} \, f_{R_\ell}(t/s)
\end{equation}
where the scaling functions $f_{C_\ell}$ and $f_{R_\ell}$ are power-laws in the long time limit, i.e.
$f_{C_\ell}(y) \sim y^{-\lambda_{C,\ell}/z}$ and $f_{R_\ell}(y) \sim y^{-\lambda_{R,\ell}/z}$ for $y \gg 1$ \cite{HenPle}.
The values of the dynamical exponent $z$ as well as of the scaling exponents $b_\ell$, $a_\ell$, $\lambda_{C,\ell}$, and 
$\lambda_{R,\ell}$ depend on the dynamical universality class. For growth processes the dynamical exponent $z$ is related
to the roughness exponent $\alpha$ and the growth exponent $\beta$ by the relation $z = \alpha/\beta$. As shown in the studies
\cite{Kal99,Rot06,Rot07,Noh10} the local autocorrelation and autoresponse exponents, $\lambda_{C,\ell}$ and
$\lambda_{R,\ell}$, are identical. In addition, they are given by $\lambda_{C,\ell}=
\lambda_{R,\ell} = d$ in systems described by a linear stochastic equation. 
Here $d$ is the dimensionality of the substrate. In addition, the scaling exponent of the correlation
function, $b_\ell$, is given by $b_\ell = - 2 \alpha/z$, whereas for the scaling exponent of the response function
the relation $a_\ell = d/z -1$ can be conjectured. As for the linear stochastic equations we have that $\alpha =
(z-d)/2$, it follows that in linear systems both scaling exponents have the same value, $a_\ell=b_\ell=d/z -1$.
This is different for the non-linear Kardar-Parisi-Zhang equation \cite{Kar86} where for a one-dimensional substrate
we have $\alpha = 1/2$ and $z = 3/2$, yielding
$b_\ell = -2/3$ and $a_\ell = -1/3$ \cite{Noh10}.

In this 
paper we study the ageing behaviour of certain global two-time quantities, namely the
correlation function of the squared width and the response of the squared width to a global perturbation.
We show that the scaling behaviour of the global response depends in linear systems on how the system is perturbed, yielding 
different results for different protocols. This observation is of interest as the global response should
be readily accessible in smoothening experiments \cite{Ngu09}, for example. Exploiting the fact that exact solutions can be obtained for growth
processes described by linear Langevin equations, we comprehensively study the scaling behaviour of these global two-time
quantities for the Edwards-Wilkinson 
\cite{Edw82} and the noisy Mullins-Herring equations \cite{Mul63}, thereby distinguishing between various limiting cases. 
In order to gain some understanding of the behaviour of these quantities in non-linear systems,
we also discuss some data that have been obtained by numerically integrating the non-linear 
Kardar-Parisi-Zhang equation \cite{Kar86}. Our results indicate that in the non-trivial correlated regime the two-time global
quantities generically exhibit a behaviour of full ageing. We also discuss 
the global fluctuation-dissipation ratio and show how this quantity can be used for the characterisation of growth processes.

Our paper is organised in the following way. In the next section we remind the reader of the typical behaviour of a
growing surface. This is done with the help of a microscopic deposition model \cite{Chou09a} that is very well
described by the one-dimensional Edwards-Wilkinson equation. This model also allows us to motivate the different
global perturbations that we are going to discuss in the following sections. Section 3 is devoted to
the exact computation of the global two-time quantities in cases that are described by linear stochastic Langevin
equations. The exact results allow us to comprehensively investigate all possible dynamical regimes. In Section 4 we present our
results for the one-dimensional Kardar-Parisi-Zhang equation where we highlight some commonalities with
and differences to the linear systems
discussed in the previous section. Finally, Section 5 gives out conclusions.

\section{Motivation: A deposition model}
In the following we briefly discuss the behaviour of a simple deposition model that turns out to be an excellent 
representative of the Edwards-Wilkinson universality class. This model is then used to motivate the different protocols 
discussed in the next sections for measuring the global response.

\subsection{Definition of the model}
Our deposition model \cite{Chou09a} is based on Family's original
random deposition with surface relaxation (RDSR) process \cite{1986FF} and differs from this model by the diffusion
step. In the RDSR process a particle deposited on the surface is allowed to
jump to one of the neighbouring sites if this site has a lower height than
the site of deposition. In our model we assign an energy $E_i(t)=g\,h_i(t)$
to the column at site $i$ where $h_i(t)$ is the height of
that column at time $t$. The constant $g$ can be thought of as the
gravitation constant, for example. 
Starting from an initially flat substrate, particles of mass one are
deposited on randomly chosen sites and then allowed to diffuse locally after
deposition. For a diffusion step taking place at time $t$, we select one of
the neighbouring sites $j$ at random and accept the jump with the
temperature and time dependent (Metropolis like) probability
\begin{equation}
P_{i\longrightarrow j}(T,t)=\left\{ 
\begin{array}{l}
1~~~\mbox{if}~~E_j(t)\leq E_i(t)\nonumber \\ 
e^{-\left[ E_j(t)-E_i(t)\right] /k_BT}=e^{-g\left[ h_j(t)
-h_i(t)\right] /k_BT}~~~\\
\hspace*{4cm} \mbox{if}~~E_j(t)>E_i(t)%
\nonumber
\end{array}
~.\right.  \label{jump_rule}
\end{equation}
In the following we choose units thus that the Boltzmann constant $k_B =1$.

In contrast to the original model there is a non-vanishing probability that
a deposited particle jumps to a neighbouring site with a higher height than
the deposition site. We assume this jump to be thermally activated and to
depend on $T$ (the temperature of the substrate). As we discuss in the
following, the ratio $T/g$ is a parameter that governs the morphology of the growing
interface and allows us to study the response of the surface to a change in
external conditions.

It is instructive to look at the behaviour of the model in the limits of $%
T\rightarrow 0$ and $T\rightarrow \infty $. At zero temperature no jumps to
sites with higher height are allowed, and the particle is incorporated into
the aggregate at the selected neighbouring site if the column at that site is
shorter than at the initial site.
Thus, for $T\rightarrow 0$ we recover the RDSR process \cite{1986FF}.
In the opposite limit, $T\rightarrow \infty $, however, a particle will
always jump to the selected neighbouring column, irrespective of the height
difference. As a result, the different columns will grow independently,
yielding an uncorrelated surface as for the RD process. For intermediate
temperatures, a crossover between the RD and the RDSR processes is observed,
with the crossover point depending on the temperature.

\begin{figure}[h]
\centerline{\epsfxsize=4.25in\ \epsfbox{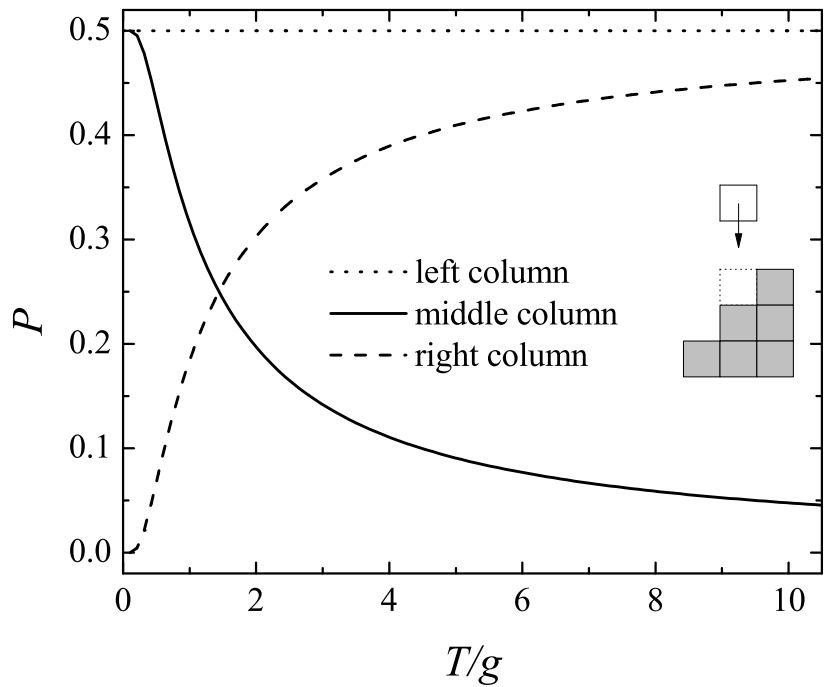}}
\caption{Probabilities that for the shown
configuration the particle, initially deposited on top of the middle column,
comes to rest on top of one of the three columns. }
\label{fig1}
\end{figure}

The temperature dependence of the model is illustrated in Figure \ref{fig1}.
Suppose a particle is deposited on top of the middle column in the
configuration shown, the plot provides the $T/g$ dependence of the
probabilities for having this particle end up at one of the three sites. In
the original Family model the particle would always come to rest on top of
the left column.

Of course, temperature has been introduced in theoretical studies of growth
processes prior to our work (see \cite{Sar96,Men96,Bar99,Liu05,Els04}
for some examples). For a typical microscopic model studied in this
context one often assumes that the atom's hopping follows an Arrhenius-like
rate that is proportional to $e^{-E_n/k_BT}$, where the activation energy $%
E_n$ is itself proportional to the number of bonds formed by the atom before
the hopping attempt. Whereas this is surely a rather realistic modelling of
diffusion processes on a crystal surface, we have opted here for a simpler
approach where the probability for a particle to hop depends on the height
difference between the actual site and the proposed new site. (Note that
jumps to sites with higher heights have also been allowed in other
microscopic growth models \cite{Sar96,Gha01}). As we will show in the
following, all aspects of our simple model agree perfectly with the solution
of the stochastic Edwards-Wilkinson (EW) equation with a {\em single} fit parameter.

\subsection{Interface width}
In the following we are interested in the
dependence of the surface width $W$ on time $t$, on the ratio $T/g$, and on
system size $L$ (in this section we only discuss the one-dimensional 
case). In our simulations, we have a wide range of $T/g$ values and
several $L$'s up to 1000. For simplicity, we always start from a flat
surface, i.e.\ $W=0$ for $t=0$. For the data discussed below, the unit of $t$
is one Monte Carlo Step (i.e., $L$ particles deposited) and we averaged over 1000
independent runs with different random numbers.

\begin{figure}[h]
\centerline{\epsfxsize=5.25in\ \epsfbox{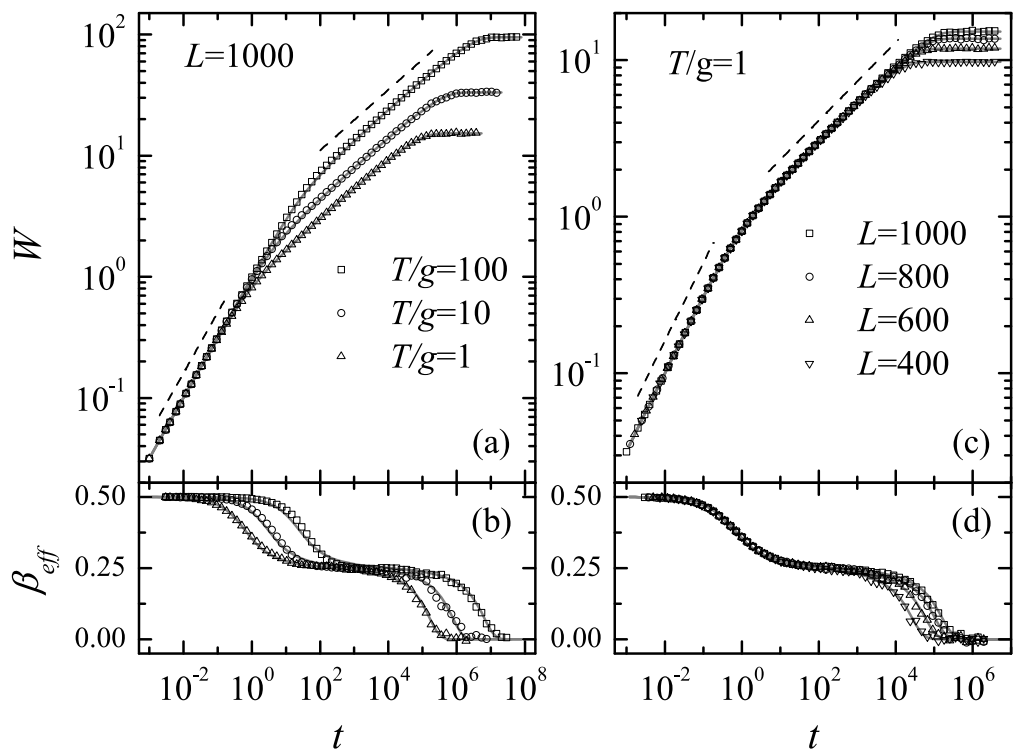}}
\caption{(a) Log-log plot of the surface width vs time for a system of size
$L=1000$ and different values of $T/g$. The dashed lines have the slopes $1/2$
and $1/4$ expected in the random deposition and EW regimes, respectively.
The location of both crossover points depend on temperature. The full lines
are obtained from fitting the exact solution of the EW stochastic equation.
Here and in the following error bars are smaller than the symbol sizes. (b)
Time evolution of the effective exponent (\ref{beff}) for the data shown in (a).
(c) Log-log plot of the surface width vs time for systems of different sizes
evolving at the value $T/g=1$. (d) Time evolution of the effective
exponent (\ref{beff}) for the data shown in (c). The full lines are
derived from the fits to the exact solution of the EW stochastic equation. }
\label{fig2}
\end{figure}

In Figure \ref{fig2}a,c we show the time dependence of the surface width for,
respectively, the case with fixed $L=1000$ at different $T/g$'s and the case
with a fixed $T/g=1$ and various $L$'s. As for the RDSR process one
distinguishes three regimes separated by two crossover points: a random
deposition (RD) regime, followed by a EW regime, with a final crossover to
the saturation regime. In contrast to Family's original model, the initial RD
process is not confined to very early times
$t\leq 1$ but extends to larger times. In fact, the crossover time $%
t_1$ between the RD and the EW regimes is shifted to higher values for
increasing temperatures and diverges in the limit of infinite temperatures.
As the crossover is smeared out, we identify the
crossover point with the intersection point of the straight lines fitted to
the two linear regimes in the log-log plots. Due to the nature of the
uncorrelated deposition of particles (Poisson process), we have the identity
$W^2=t$ in the RD regime, yielding the width $W_1=\sqrt{t_1}$ at the
crossover point. In the EW regime the relation between width and deposition
time changes to $W\propto t^{1/4}$. This regime extends up to a second
crossover point $(t_2,W_2)$, whose precise location also depends on the
value of $T$ and beyond which the final saturation regime prevails.
The crossover
between the different regimes is further illustrated in Fig. \ref{fig2}b,d
where we show the time evolution of the effective exponent
\begin{equation}
\beta _{eff}=\frac{d\ln W}{d\ln t}  \label{beff}
\end{equation}
for the two cases.

Next, we turn our attention to the stochastic EW equation (see also the next section)
\begin{equation}
\frac{\partial h({\bf x},t)}{\partial t}=\nu \nabla ^2h({\bf x},t)+\eta
({\bf x},t)  \label{eqEW}
\end{equation}
where $\eta ({\bf x},t)$ is a Gaussian white noise with zero mean and
covariance $\left\langle \eta ({\bf x},t)\eta ({\bf y},s)\right\rangle
=D\delta ^d({\bf x}-{\bf y})\delta (t-s)$ ($d$ is the dimensionality of the substrate) and $\nu$ is the surface tension
or diffusion constant.
This equation can be solved exactly to give us in one dimension the width\ (squared)
\begin{equation}
W^2\left( t\right) =\frac D{2\nu L}\sum_n\frac{1-e^{-2\nu tq_n^2}}{q_n^2}
\label{eqWsq}
\end{equation}
where $q_n=2\pi n/L$ and the sum is over $\left[ -L/2,L/2\right] $ but {\em %
excluding} the zero mode:{\em \ }$n=0$. The effective exponent is therefore
given by

\begin{equation}
\beta _{eff}=\frac{\nu t\sum\limits_n e^{-2\nu tq_n^2}}{\sum\limits_n \left[ 1-e^{-2\nu
tq_n^2}\right] /q_n^2}~.  \label{eqBeta}
\end{equation}
In the expression (\ref{eqWsq}), we must fix $D=1$ in order to agree with $%
W^2=t$ in the RD regime. This leaves us with just one free parameter,
namely, $\nu \left( T/g\right) $, which can be obtained by fitting the numerical data
to the theoretical expression, see Table \ref{table1}. The result of this procedure
can be summarized by
\begin{equation}
\nu =\frac 1{a+b\,T/g}  \label{nuT}
\end{equation}
with $a=4.23$ and $b=2.13$. As the solid lines in Fig. \ref{fig2} show, excellent fits are
achieved with these values of $\nu$.
Clearly, the theoretical curves are in very good
agreement with the data over the entire range of $L$'s and $T/g$'s explored.
Note in addition that it is possible to collapse {\em all} data onto a single curve \cite{Chou09a}.

It is worth mentioning that a $1/T$ dependence of the surface tension $\nu$ is typically encountered in
experiments on step fluctuations \cite{Bar96,Dou02,Dou05,Bon05} or
on surface smoothening \cite{Ngu09} which are described theoretically by linear Langevin equations.

\begin{table}[tbp]
\begin{center}
\begin{tabular}{|c|c|c|c|c|}
\hline
$T/g$ & $0.1$ & 1 & 10 & 100 \\ \hline
$\nu$ & 0.2300 & 0.1768 & 0.0379 & 0.0046 \\ \hline
\end{tabular}
\end{center}
\caption{Values for the surface tension $\nu$ at different values of $T/g$ that
result from a fit of the numerical data to the exact solution of the
Edwards-Wilkinson equation.}
\label{table1}
\end{table}

The evolution of the width shown in
Fig. \ref{fig2} is very reminiscent of the behaviour encountered in
competitive growth models
\cite{Hor01,Cha02,Hor03,Kol04,Mur04,Bra05,Iru05,Hor06,Rei06, Oli06,Kol06}. In
both cases two different scaling regimes are found whose ranges depend on a
system parameter. In the competitive growth models one considers a mixture
of two different deposition processes where one of them takes place with
probability $p$ whereas the other takes place with probability $1-p$. One
example is the RD/RDSR model \cite{Hor01} where the deposition happens
according to the RDSR rules with probability $p$ and to the RD rules with
probability $1-p$. Whereas for $p=1$ and $p=0$ only one of the processes is
realized, for general values of $p$ the mixture of the two processes leads
to a crossover between the two regimes where the crossover time and width
depend on the value of $p$. In our model, the ratio $T/g$ plays the role of the
quantity $p$, the main difference being that we do not artificially choose
between the different processes, as in our system the competition is
intrinsic and governed by the value of the temperature.

\subsection{Global perturbations of the growth process}
Experimentally, a quantity that can be changed easily and that gives way to a global
perturbation is the temperature. If an experimental system is described by a stochastic
Langevin equation, temperature can enter either through the surface tension $\nu$
(also called diffusion constant or mobility, depending on the physical context) or through the
noise. In the next sections we shall study the responses to two different global perturbations:
either we keep the noise unchanged and suddenly change $\nu$ (as it is the case in 
surface smoothening experiments or in step fluctuation studies) or we keep the surface tension constant and
change the noise (as it is the case in some deposition experiments where the noise in the particle
flux can be changed experimentally). Both protocols have been used for studying the
change in morphology of a growing surface when changing experimental conditions \cite{Maj96,Chou09b}.

We can use our deposition model in order to see which
global quantity is best suited to study the changes due to such a global perturbation.
Let us consider at temperature $T$ a configuration where the height of the deposed column at site $i$
is $h_i$. Setting the lattice constant in vertical direction to be 1 and setting the
potential energy of the initial flat surface to be zero, the potential energy $U_i$ stored
in a column of height $h_i$ is (with the mass of a deposed particle set to 1)
\begin{equation}
\frac{U_i}{T} = \frac{g}{T} + 2 \frac{g}{T} + \cdots \ + (h_i -1 ) \frac{g}{T} = h_i(h_i-1) \frac{g}{2 T}~.
\end{equation}
Shifting the value zero of the potential energy to the average height $\overline{h} =
\frac{1}{N} \sum\limits_{i=1}^N h_i$, where $N$ is the number of sites on the substrate,
we obtain for the total potential energy of our
configuration the value
\begin{equation}
\frac{U}{T} = \sum\limits_{i=1}^N \frac{U_i}{T} = \frac{g}{2 T} \sum\limits_{i=1}^N  \left( h_i(h_i-1) - \overline{h} (\overline{h}-1) \right) 
= \frac{g}{2 T} \sum\limits_{i=1}^N \left( h_i^2 - \overline{h}^{\, 2} \right) = \frac{g N}{2 T} W^2
\end{equation}
where $W^2 = \frac{1}{N} \sum\limits_{i=1}^N \left( h_i^2 - \overline{h}^{\, 2} \right) 
= \frac{1}{N} \sum\limits_{i=1}^N \left( h_i - \overline{h} \right)^2$ is the squared
width. From this it follows that $\frac{N}{2} W^2$ is the quantity conjugated to  $g/T$ and, as $g/T \sim \nu$, to $\nu$.
This motivates the use of the square of the surface width (instead of the surface width itself) as the global quantity
to be studied in the following.

\section{Linear Langevin equations}
Linear Langevin equations \cite{Kru97} are discussed in a variety of physical situations, ranging from 
equilibrium step fluctuations \cite{Gie01,Dou04,Bon05,Dou05} to film growth \cite{Wol90,Gol91,Das91}
and from magnetic systems \cite{HenPle} to elastic lines in a random environment \cite{Igu09}.
In the following we focus on two simple but generic cases which can be summarised by the equation
\begin{equation}
\frac{\partial h(\mathbf{x},t)}{\partial t}=-\nu(i\nabla)^m h(\mathbf{x},t)+\eta(\mathbf{x},t),
\label{eqLa}
\end{equation}
where $m$ is an even number and the noise is uncorrelated and with zero mean,
\begin{equation}
\langle \eta(\mathbf{x},t) \eta(\mathbf{x}',t')\rangle=D \delta^d(\mathbf{x}-\mathbf{x}') \delta(t-t')~.
\label{eqC2eta}
\end{equation}
For $m=2$ we recover the well-known Edwards-Wilkinson (EW) equation \cite{Edw82} and for $m=4$ 
we obtain the noisy Mullins-Herring (MH) equation \cite{Mul63}. Note that for these systems the dynamical exponent
$z=m$.

As already discussed in the introduction, some aspects of ageing in these systems have been elucidated
previously through the study of local quantities. 
In the following we are going to characterise this relaxation through 
global two-times quantities related to the square of the surface width.

\subsection{Global two-time correlation function}
The first global quantity that we are going to discuss is the connected correlation function
(here and in the following $\langle \cdots \rangle$ indicates an average over the noise)
\begin{equation}
C(t,s) = \langle \widetilde{W}^2(t) \, \widetilde{W}^2(s) \rangle - \langle \widetilde{W}^2(t)\rangle\langle \widetilde{W}^2(s)\rangle
\label{eq:corr}
\end{equation}
where 
\begin{equation}
\widetilde{W}^2(t) = \frac{L^d}{2} W^2(t) = \frac{1}{2} \sum\limits_{\mathbf{x}}  \left( h(\mathbf{x},t)  - \overline{h} \right)^2
\label{eq:W2tilde}
\end{equation}
and
\begin{equation}
W^2(t) = \frac{1}{L^d} \sum\limits_{\mathbf{x}}  \left( h(\mathbf{x},t)  - \overline{h} \right)^2
\label{eq:W2}
\end{equation}
is the square of the surface width 
at time $t$ on top of a $d$-dimensional substrate of linear extension $L$. Here $\mathbf{x}$ labels the
different substrate sites.
We assume in the following that $t > s$ and call $s$ the waiting time and $t$ the observation time.

Writing both the height $h(\mathbf{x},t)$ and the noise $\eta(\mathbf{x},t)$ as a sum over reciprocal lattice
vectors,
\begin{equation}
h(\mathbf{x},t)=\sum_{\mathbf{q}}h_\mathbf{q}(t)\exp(i\mathbf{q}\cdot\mathbf{x}) ~~, ~~
\eta(\mathbf{x},t)=\sum_{\mathbf{q}}\eta_\mathbf{q}(t)\exp(i\mathbf{q}\cdot\mathbf{x})~,
\end{equation}
equations (\ref{eqLa}) and (\ref{eqC2eta}) become
\begin{equation}
\frac{\partial h_{\mathbf{q}}}{\partial t}=-\nu\mathbf{q}^{m} h_\mathbf{q}+\eta_\mathbf{q}
\label{eqFTLa}
\end{equation}
and
\begin{equation}
\langle \eta_\mathbf{q}(t) \eta_\mathbf{q}'(t')\rangle=D L^{-d} \delta^d_{\mathbf{q}+\mathbf{q}'} \delta(t-t')~.
\end{equation}
The general solution of (\ref{eqFTLa}) is readily obtained to be
\begin{equation}
h_{\mathbf{q}}(t)=\exp(-\nu\mathbf{q}^{m}t)h_{\mathbf{q}}(0)+\int_0^t d\tau\exp
\left(-\nu\mathbf{q}^{m}(t-\tau)\right)\eta_\mathbf{q}(\tau)
\label{eqSFTLa}.
\end{equation}
Assuming an initially flat surface, $h_{\mathbf{q}}(0) = 0$ for every $\mathbf{q}$, one obtains the following two-point 
correlation function of the surface height in $\mathbf{q}$-space:
\begin{equation}
\langle h_{\mathbf{q}_1}(t_1) h_{\mathbf{q}_2}(t_2) \rangle = \frac{D}{L^d\nu}e^{-\nu(q_1^m t_1+q_2^m t_2)}\frac{1}{q_1^m+q_2^m}\left(e^{\nu(q_1^m+q_2^m)t_{<}}-1\right) \delta^d_{\mathbf{q}_1+\mathbf{q}_2},
\label{eqC2FTh}
\end{equation}
where $q_i=|\mathbf{q}_i| $ and $t_<$ is the smaller of the times $t_1$ and $t_2$.
This well-known result will be useful in the following.

For the computation of the correlation function (\ref{eq:corr}) we remark that the square of the surface width (\ref{eq:W2})
can be written in terms of the Fourier components of the height as
\begin{equation}
\widetilde{W}^2(t)=\frac{1}{2}\sum_{\mathbf{x}}\left(\sum_{\mathbf{q}\neq 0}h_{\mathbf{q}}(t)\exp{(i\mathbf{q}\cdot\mathbf{x})}\right)^2
\label{eqW}
\end{equation}
where we took into account that the average height of the surface is equal to $h_{\mathbf{q=0}}$.
This then yields
\begin{eqnarray}
\langle \widetilde{W}^2(t)\widetilde{W}^2(s)\rangle & = & \frac{1}{4} \sum_{\mathbf{x},\mathbf{x}'}\sum_{\mathbf{q},\mathbf{q}',\mathbf{p},\mathbf{p}'\neq 0}
\langle h_{\mathbf{q}}(t)h_{\mathbf{q}'}(t)h_{\mathbf{p}}(s)h_{\mathbf{p}'}(s)\rangle \nonumber \\
&& \exp{[i(\mathbf{q}+\mathbf{q}')\cdot\mathbf{x}]}\exp{[i(\mathbf{p}+\mathbf{p}')\cdot\mathbf{x}']}
\label{eqC4W1}.
\end{eqnarray}
The calculation of (\ref{eq:corr}) is therefore reduced to the determination of the four-point
correlation function of surface height in $\mathbf{q}$-space. Using the Gaussian statistics of the linear equations, this four-point
function can be expressed through two-point functions of the form (\ref{eqC2FTh}), yielding
\begin{eqnarray}
\langle h_{\mathbf{q}_1}(t_1) h_{\mathbf{q}_2}(t_2) h_{\mathbf{q}_3}(t_3) h_{\mathbf{q}_4}(t_4) \rangle & = & \langle h_{\mathbf{q}_1}(t_1) h_{\mathbf{q}_2}(t_2) \rangle\langle h_{\mathbf{q}_3}(t_3) h_{\mathbf{q}_4}(t_4) \rangle\nonumber\\
& &+\langle h_{\mathbf{q}_1}(t_1) h_{\mathbf{q}_3}(t_3) \rangle\langle h_{\mathbf{q}_2}(t_2) h_{\mathbf{q}_4}(t_4) \rangle\nonumber\\
& &+\langle h_{\mathbf{q}_1}(t_1) h_{\mathbf{q}_4}(t_4) \rangle\langle h_{\mathbf{q}_2}(t_2) h_{\mathbf{q}_3}(t_3) \rangle
\label{eqC4FTh1}.
\end{eqnarray}
Inserting this into (\ref{eqC4W1}) and using the identity 
\begin{equation}
\sum_{n=0}^{N-1}\exp\left(\frac{2\pi i n m}{N}\right)=N\delta_{m,0}
\end{equation}
yields the expression
\begin{eqnarray}
\langle \widetilde{W}^2(t)\widetilde{W}^2(s)\rangle & = & \frac{D^2}{4\nu^2}\left\{\sum_{\mathbf{q},\mathbf{p}\neq 0}\frac{1}{4q^mp^m}
\left(1-e^{-2 q^m \nu t}\right)\left(1-e^{-2 p^m \nu s}\right) \right. \nonumber \\
 && + \left. \sum_{\mathbf{q}\neq 0}\frac{1}{2q^{2m}}e^{-2 q^m \nu t}\left(e^{2 q^m \nu s}+e^{-2\nu q^m s}-2\right)\right\}
\nonumber
\label{eqC4W2}.
\end{eqnarray}
The first term on the right-hand side being just $\langle \widetilde{W}^2(t)\rangle\langle \widetilde{W}^2(s)\rangle$, we finally obtain
\begin{equation}
C(t,s)=\frac{D^2}{8\nu^2}\sum_{\mathbf{q}\neq 0}\frac{1}{q^{2m}}e^{-2 q^m \nu t}\left(e^{2 q^m \nu s}+e^{-2 q^m \nu s}-2\right).
\label{eqC4W4}
\end{equation}

The behaviour of $C(t,s)$ is controlled by two length scales: $l_t \equiv (2\nu t)^{1/m}$ and $l_s \equiv (2\nu s)^{1/m}$,
with $l_t > l_s$. 
Depending on the values of $l_t$ and $l_s$ and their relations to the maximum
and minimum values of $q$, $q_{max} = \pi \sqrt{d}$ and $q_{min} = 2 \pi/L$, different regimes are encountered. For example,
if at time $t$ the length $l_t < 1/q_{max}$\footnote{By $l_t < 1/q_{max}$ we mean that the order of $l_t$ has
to be much smaller than the order of $1/q_{max}$. There are complicated crossover effects showing up when
$l_t$ and $1/q_{max}$ are of comparable magnitude that we are not going to discuss. Of course, the same remark
applies when comparing one of the length scales to $1/q_{min}$.}, one still is in the short time regime where the random deposition
process prevails. On the other hand, if at time $t$ one has that $l_t> 1/q_{min}$, one is the saturation regime.
Finally, the system is in the correlated regime when $1/q_{max}<l_t<1/q_{min}$. 
We discuss in the following the functional form of the correlation function for the different regimes (see
Table \ref{table2} for a summary).

1. If $l_t <1/q_{max}$ we are in the short time regime. In that case
all exponentials in (\ref{eqC4W4}) can be expanded and only the leading terms need to be retained, i.e.
$e^{-2q^m\nu t} \approx 1-2q^m\nu t$ and $e^{2q^m\nu s}+e^{-2q^m\nu s}-2\approx \left(2 q^m \nu s \right)^2$ for
all $q$. Replacing the sum by an integral, we obtain
\begin{eqnarray}
C(t,s)
&\approx&
\frac{D^2s^2}{2}\left(\frac{L}{2\pi}\right)^d\Omega_d
\int_0^\pi q^{d-1}\left(1-2q^m\nu t\right)dq\nonumber\\
&\approx& \frac{D^2\pi^{d/2}L^d}{2^d\Gamma(d/2)d}s^2\left(1-\frac{2d}{d+m}\pi^m\nu t \right),
\label{eqC4W7}
\end{eqnarray}
where $\Omega_d=2\pi^{d/2}/\Gamma(d/2)$ is the solid angle of the $d$-dimensional sphere.
Therefore, in the short time regime the correlation function decreases linearly with time.

\begin{figure}[h]
\centerline{\epsfxsize=5.25in\ \epsfbox{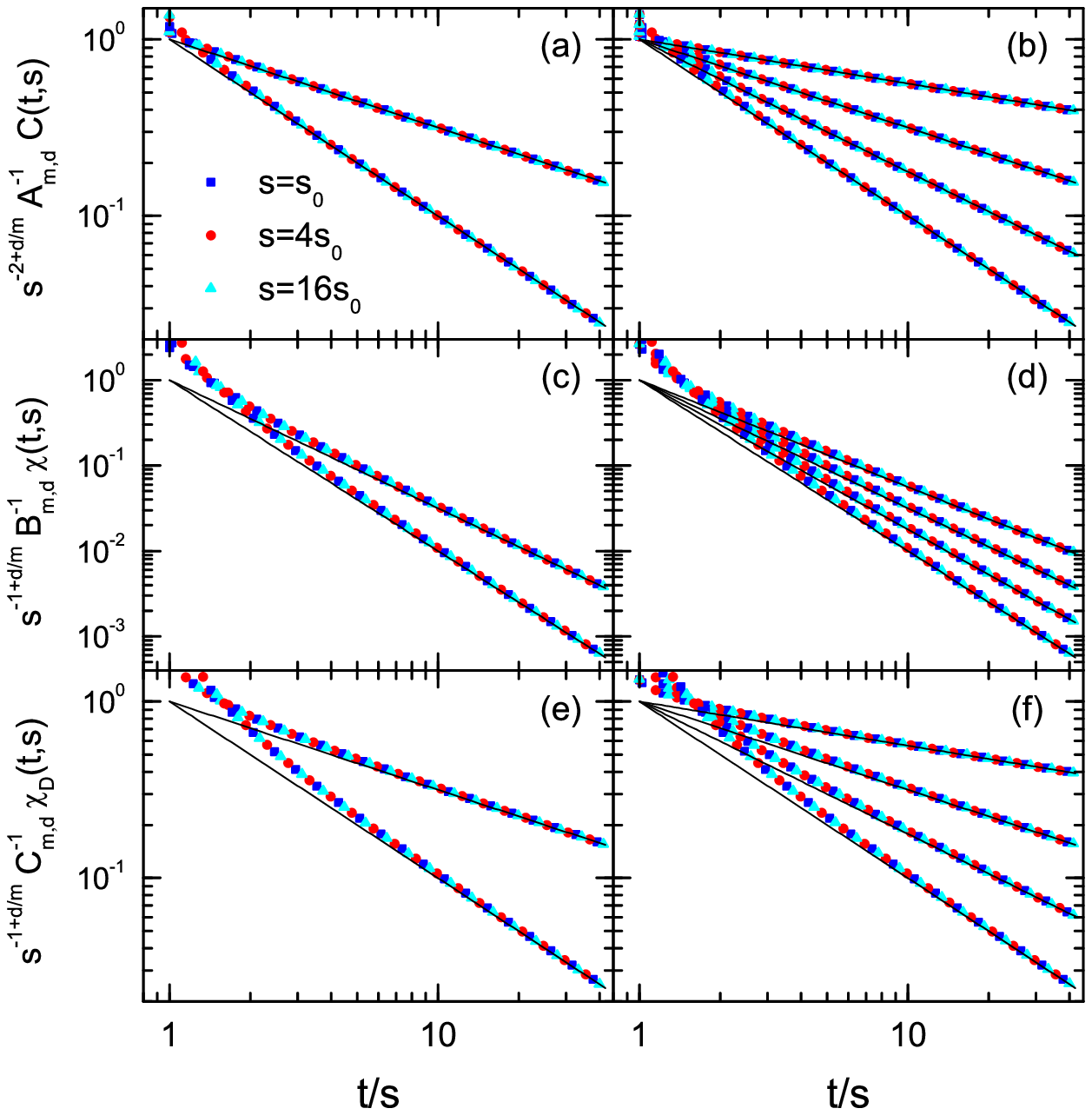}}
\caption{Global correlation (a,b), global response to a change of the surface tension (c,d), and
global response to a change of the noise strength (e,f) when both times $t$ and $s$ are in the
correlated regime. The symbols are obtained by numerically evaluating the exact expressions 
(\ref{eqC4W4}), (\ref{eqchi1}), and (\ref{eqchiD2}) whereas the lines indicate the asymptotic power-laws 
(\ref{eqC4W9}), (\ref{eqchi3}), and (\ref{eqchiD3}).
Panels (a), (c), and (e) show data for the EW case and dimensions $d=1, 2$ (from top to bottom), panels (b), (d) and (f)
show data for the MH  case and dimensions $d=1, 2, 3, 4$ (from top to bottom). The different symbols correspond
to different waiting times where $s_0 = 1000$ below the critical dimension and $s_0 = 200$ at the critical
dimension (which is $d=2$ for EW and $d=4$ for MH). The linear extension of the system is $L= 2^{14}$ 
for $d=1$, $L= 2^9$ for $d=2$, $L=2^7$ for $d=3$, and $L=2^6$ for $d=4$. 
}
\label{fig3}
\end{figure}

2. If the final time is in the correlated regime, i.e. if $1/q_{max}<l_t<1/q_{min}$, one has to distinguish between two
different cases, depending on whether $l_s < 1/q_{max}$ (the system was in the RD regime at time $s$) or
$1/q_{max} < l_s < l_t$ (the system was in the correlated regime at time $s$). 
In the first case we can again replace $e^{2q^m\nu s}+e^{-2q^m\nu s}-2$ by $\left(2 q^m \nu s \right)^2$ 
and obtain the expression
\begin{eqnarray}
C(t,s)&\approx& \frac{D^2s^2}{2}\left(\frac{L}{2\pi}\right)^d\Omega_d\int_0^\infty q^{d-1}e^{-2q^m\nu t}dq\nonumber\\
&=&A_{m,d} s^2t^{-d/m},
\label{eqC4W6}
\end{eqnarray}
where
\begin{equation}
A_{m,d}=\frac{D^2 \Gamma(d/m) L^d}{2^{d+d/m}\pi^{d/2}m\nu^{d/m}\Gamma(d/2)}~,
\label{eqA}
\end{equation}
yielding a power-law decay with an exponent $d/m$.
In the second case, which is the most interesting one as both times are in the correlated regime,
we expand all exponentials in (\ref{eqC4W4}) and integrate term by term. This yields
\begin{eqnarray}
C(t,s) &=& A_{m,d} s^2\left\{t^{-d/m}+\frac{s^2}{12}\frac{\partial^2}{\partial t^2}t^{-d/m}+\frac{s^4}{360}\frac{\partial^4}{\partial t^4}t^{-d/m}+\cdots \right\}\nonumber\\
&=& A_{m,d} s^2 t^{-d/m}
\left\{1+\frac{1}{12}\frac{d}{m}\left(\frac{d}{m}+1\right)
\left(\frac{s}{t}\right)^2+ \right. \nonumber \\
&& \left. \frac{1}{360}\frac{d}{m}\left(\frac{d}{m}+1\right)
\left(\frac{d}{m}+2\right)\left(\frac{d}{m}+3\right)\left(\frac{s}{t}\right)^4
+\cdots\right\}\nonumber\\
&\approx& A_{m,d} s^2 t^{-d/m}
\label{eqC4W9}
\end{eqnarray}
where the last line gives the asymptotic form for $t \gg s$. Interestingly, we recover asymptotically the same
leading behaviour as for the case $l_s < 1/q_{max}$, i.e. in the long time limit any memory of the regime that
prevailed at time $s$ is lost.
Looking closer at the expressions (\ref{eqC4W6}) and (\ref{eqC4W9}), we see that in both cases the global correlation
function can be cast in the standard form \cite{HenPle}  $C(t,s) = s^{-b} f_C(t/s)$ where $b = d/m-2$ and $f_C(t/s)$ is a scaling function
that only depends on the ratio $t/s$. In addition, the scaling function is asymptotically given by a power-law,
$f_C(t/s) \sim (t/s)^{- \lambda_C/z}$,
with the autocorrelation exponent $\lambda_C =  d$ and the dynamical exponent $z = m$. 
In Figure \ref{fig3}a,b we compare the correlation function
(\ref{eqC4W4}), where the sum has been evaluated numerically, with the 
asymptotic power-law (\ref{eqC4W9}), see Figure \ref{fig3}a for the EW case and Figure \ref{fig3}b for the MH case. 
In all cases the asymptotic regime is accessed very rapidly.

We can now also compare for this regime the global correlation function with the local height-height correlation,
see \cite{Kal99,Rot06,Rot07}. For the local correlation function one obtains the scaling exponents 
$b_\ell = d/m-1$ and $\lambda_{C,\ell} = d$.
Thus, for both the local and global quantities the long-time behaviour is governed by the same power-law exponent, 
$\lambda_C = \lambda_{C,\ell}$,
but the value of the scaling exponent $b$ is different from the value of $b_\ell$. This yields interesting differences
at the upper critical dimension which is $d=2$ for EW and $d=4$ for MH.
Indeed, whereas the local exponent $b_\ell$ then vanishes and the scaling function of the local correlation shows a
logarithmic dependence on time
(which makes 
the extraction of the correct scaling behaviour from experimental or simulation data rather tedious) \cite{Rot06}, 
the global exponent remains 
different from zero and no logarithms are occulting the scaling behaviour of the global correlation, 
making this a much better quantity for studying dynamical scaling close
to the upper critical dimension.

\begin{table}[tbp]
\begin{center} 
\begin{tabular}{|c||c|c|c|}
\hline
$s$ \textbackslash ~ $t$ & RD& CR & SR \\ \hline\hline
RD & $a_1 \, s^2 \, (1 - c_1 \, t)$ &  $a_2  s^{2-d/m} \, (t/s)^{-d/m}$ & $a_3 \, s^2 \, e^{-b t}$  \\ \hline
CR & $-$ & $a_2  s^{2-d/m} \, (t/s)^{-d/m}$ &  $a_3 \, s^2 \, e^{-b t}$\\ \hline
SR & $-$ &  $-$ &  $\tilde{a}_3 e^{-b (t-s)}$ \\ \hline
\end{tabular}
\end{center}
\caption{Summary of the asymptotical dependence of the global correlation function for the different cases.
RD corresponds to the random deposition regime, CR to the correlated regime, and SR to the saturation regime.
The values of the different coefficients can be read off from the equations
in the main text.
}
\label{table2}
\end{table}

3. In case the observation time is in the saturation regime with $l_t > 1/q_{min}$, the leading contribution
is given by the term in $e^{-2q_{min}^m\nu t}$. If the waiting time is also in the saturation regime, we have
$l_s > 1/q_{min}$ and therefore
\begin{equation}
C(t,s)\approx \frac{dD^2L^{2m}}{4(2\pi)^{2m}\nu^2}e^{-2 q_{min}^m \nu (t-s)},
\label{eqC4W10}
\end{equation}
i.e. the global correlation is time-translation invariant. This is of course expected, as the saturation regime
corresponds to the steady state regime. If the waiting time is such that $l_s < 1/q_{min}$, then 
we can approximate for $q = q_{min}$ the term involving $s$ by $\left(2 q_{min}^m \nu s \right)^2$ which yields
\begin{equation}
C(t,s)\approx d D^2s^2e^{-2q_{min}^m\nu t}~.
\label{eqC4W5}
\end{equation}
It is worth noting that in both cases the decay time in the exponential, $\tau = 1/(2 q_{min}^m \nu)$, depends on
the diffusion mechanism through the values of $\nu$ and $m$ but does not depend on the dimensionality of the substrate.

\subsection{Global two-time response functions}
In order to measure the global response
we are going to perturb the growing system either through a change in the surface tension $\nu$ or in the strength
$D$ of the noise correlator, as discussed in Section 2.3.
Having thus perturbed our system,
we then study its relaxation to the state that is realised when the system
evolves all the time at fixed values of $\nu$ and $D$. 

In the first protocol, we start from an
initially flat surface and let the system evolve with a surface tension $\mu$ until the waiting time $s$
after which we set the surface tension to the final value $\nu$.
In order to monitor the relaxation of the system for $t > s$, we compute the function
\begin{equation}
\chi(t,s)=\frac{\langle \widetilde{W}^2 \rangle_{\mu\rightarrow \nu}(t,s)- \langle \widetilde{W}^2 \rangle_{\nu}(t)}{\epsilon}
\label{eqchi}
\end{equation}
where the index "$\mu\rightarrow \nu$" indicates the perturbed system, whereas the index "$\nu$" stands for the system 
that is evolving at a fixed value $\nu$ of the surface tension. The quantity $\epsilon\equiv \nu-\mu$
measures the strength of the perturbation. Usually one is interested in small perturbations for which $| \epsilon|  \ll 1$, and 
we will start with this in the following. However, as we discuss later, the response (\ref{eqchi}) displays 
simple ageing and dynamical scaling even for very strong perturbations. 

This protocol has been used previously in \cite{Chou09b} in order to study the asymptotic long-time behaviour
of a one-dimensional perturbed EW system. No attempt was made in that work to determine the scaling functions
in the ageing regime.

It has to be noted that the response $\chi(t,s)$ is a time integrated response that gives the reaction of the
system to a perturbation lasting until time $s$. This time integrated response is related to the response $R$ due
to an instantaneous perturbation by $\chi(t,s) = \int\limits_0^s \, du\, R(t,u)$. Assuming a standard scaling behaviour
for $R$, i.e. $R(t,s) = s^{-1-a} f_R(t/s)$, it follows that the time integrated response should scale as 
$\chi(t,s) = s^{-a} f_\chi(t/s)$ \cite{HenPle}.

The expression for $\chi(t,s)$ is readily obtained by remarking that the solution of the Langevin equation 
becomes for $t \geq s$
\begin{equation}
h_{\mathbf{q}}(t)=e^{-\nu\mathbf{q}^{m}(t-s)}h_{\mathbf{q},\mu}(s)+\int_s^t d\tau e^{-\nu\mathbf{q}^{m}(t-\tau)}\eta_\mathbf{q}(\tau)
\label{eqSFTLa2},
\end{equation}
where
\begin{equation}
h_{\mathbf{q},\mu}(s)=\int_0^s d\tau e^{-\mu\mathbf{q}^{m}(s-\tau)}\eta_\mathbf{q}(\tau)
\end{equation}
is the solution of a surface that, starting from a flat initial state, evolves until time $s$ at the constant surface 
tension $\mu$.
Inserting this expression into (\ref{eqW}) yields
\begin{eqnarray}
\langle \widetilde{W}^2\rangle_{\mu\rightarrow \nu}(t,s)& =& \frac{D}{4}\sum_{\mathbf{q}\neq 0}\frac{1}{q^m}\left\{  \frac{1}{\mu}e^{-2q^m\nu(t-s)}-\frac{1}{\mu}e^{-2q^m[\nu(t-s)+\mu s]} \right. \nonumber \\
&& \left. \hspace*{2.3cm} + \frac{1}{\nu}\left(1-e^{-2q^m\nu(t-s)}\right)\right\}
\end{eqnarray} 
for $t\ge s$, which finally gives the expression
\begin{equation}
\chi(t,s)= \frac{D}{4\epsilon}\sum_{\mathbf{q}\neq 0}\frac{1}{q^m}e^{-2q^m\nu(t-s)}\left\{\frac{1}{\mu}\left(1-e^{-2q^m\mu s}\right)-\frac{1}{\nu}\left(1-e^{-2q^m\nu s}\right)\right\}
\label{eqchi1}
\end{equation}
for our response function.

For small perturbations with $| \epsilon|  \ll 1$ we can proceed  as for the correlation function in the previous section.
As we have the same cases and therefore can use the same approximation schemes, we shall only quote the final expressions,
see Table \ref{table3} for a summary.

1. At early times, where $t$ is still in the RD regime, we have a linear decay,
\begin{eqnarray}
\chi(t,s)& \approx & \frac{D\pi^{d/2+m}L^d}{2^{d}\Gamma(d/2)(d+m)}s^2  \nonumber \\
&& \hspace*{1cm} \times \left\{1-2\frac{d+m}{d+2m}\pi^m 
\left[\nu(t-s)+\frac{1}{3}(\nu+\mu)s\right]+\mathcal{O}(\nu t)^2\right\}~.
\label{eqchi2} 
\end{eqnarray}

2. When $t$ is in the correlated regime, we obtain for a perturbation that lasted only until some time 
$s$ in the RD regime the following 
expression for the response:
\begin{equation}
\chi(t,s) \approx  B_{m,d} s^2 (t-s)^{-d/m-1},
\label{eqchi3a}
\end{equation}
with
\begin{equation}
B_{m,d}=\frac{D d \, \Gamma(d/m)L^d}{2^{1+d+d/m}\pi^{d/2}m^{2}\, \Gamma(d/2)\nu^{d/m+1}} ~.
\label{eqB}
\end{equation}
If, however, the perturbation ended inside the correlated regime, we have 
\begin{eqnarray}
\chi(t,s) &=& B_{m,d} s^2(t-s)^{-\frac{d}{m}-1}\nonumber\\
& & \times\left\{1-\frac{\nu+\mu}{3\nu}\left(\frac{d}{m}+1\right)
\left(\frac{t}{s}-1\right)^{-1}+\cdots\right\}\nonumber\\
&\approx& B_{m,d} s^2 (t-s)^{-\frac{d}{m}-1},
\label{eqchi3}
\end{eqnarray}
yielding the same leading behaviour as for a perturbation that already stopped in the RD regime.
In the correlated regime we can therefore cast the response in the standard form
$\chi(t,s) = s^{-a} \, f_\chi(t/s)$ with $a=\frac{d}{m}-1$ and $f_\chi(t/s) \sim (t/s)^{-\lambda_R/z}$ for
$t/s$ large, with $\lambda_R = d+m$. We show this scaling behaviour in Figure \ref{fig3} c,d
for the EW and for the MH case and compare the scaling function to the asymptotic power-law.

One can now also compare the values of the exponents with those obtained for the global correlation function
(\ref{eq:corr}) as well as with those obtained for the response of the height to a local perturbation, see \cite{Rot06}.
Comparing with the global correlation, one notes that we have the interesting situation that $a \neq b$ and $\lambda_R 
\neq \lambda_C$. It is rather uncommon that the response of some quantity and the correlation of the same quantity display
a different asymptotic behaviour in the ageing regime \cite{HenPle}. Mathematically, this intriguing observation
can be traced back to the fact that in the 
expression (\ref{eqC4W4}) for the correlation $q^{2m}$ shows up in the denominator whereas $q^m$ is encountered in the denominator of
the corresponding expression (\ref{eqchi1}) for the response function. In order to compare the global response with the local response,
we need to look at the time integrated 
local response. Inserting the expressions given in \cite{Rot06} for the local response $R_\ell$ into the integral $\chi_\ell(t,s)
= \int\limits_0^s \, du\, R_\ell(t,u)$ yields for the local time integrated response the form $\chi_\ell(t,s) = s^{-a_\ell} \,
f_{\chi_\ell}(t,s)$ where the exponent $a_\ell$ has the value $a_\ell=\frac{d}{m}-1$, whereas the exponent governing the scaling function
in the long time regime is $\lambda_{R,\ell} = d$. We therefore have that 
the scaling exponent $a = a_\ell$ is the same for both responses but that
the global and the local autoresponses differ, $\lambda_R = \lambda_{R,\ell} + m$.

3. Finally, if the observation time is in the saturation regime, we obtain that
\begin{equation}
\chi(t,s)\approx\frac{D L^{m}}{2^{m-d+2}\pi^{m}\mu \nu}e^{-2q_{min}^m\nu(t-s)}
\label{eqchi6}
\end{equation}
for a perturbation that lasts until the saturation regime and
\begin{equation}
\chi(t,s)\approx\frac{2^{d+m-1}D\pi^m}{L^{m}}s^2e^{-2q_{min}^m\nu(t-s)}
\label{eqchi4}
\end{equation}
for a perturbation that comes to an end before entering the saturation regime.

\begin{table}[tbp]
\begin{center} 
\begin{tabular}{|c||c|c|c|}
\hline
$s$ \textbackslash ~ $t$ & RD& CR & SR \\ \hline\hline
RD & $a_4 \, s^2 \, (1 - c_2 \, t - d_1 \, s )$ &  $a_5  s^{1-d/m} \, (t/s)^{-d/m-1}$ & $a_6 \, s^2 \, e^{-b (t-s)}$  \\ \hline
CR & $-$ & $a_5  s^{1-d/m} \, (t/s)^{-d/m-1}$ &  $a_6 \, s^2 \, e^{-b (t-s)}$\\ \hline
SR & $-$ &  $-$ &  $\tilde{a}_6 e^{-b (t-s)}$ \\ \hline
\end{tabular}
\end{center}
\caption{The same as Table \ref{table2}, but now for the response (\ref{eqchi1}) to a small perturbation during
which the surface tension is changed.
The values of the different coefficients can be read off from the equations
in the main text.
}
\label{table3}
\end{table}

Let us close the discussion of the first protocol by briefly looking at the case of
strong perturbations. As $\mu$ and $\nu$ then strongly differ, we need to take into account the
existence of a third length scale $l_3 = (2\mu s)^{1/m}$. This makes the analysis more involved.
We first note that for the case $\nu \gg \mu$ we always have the relation $l_t > l_s > l_3$.
It is then readily verified that we have basically the same three cases as before, with  
the response given by the expression (\ref{eqchi2}) respectively (\ref{eqchi3}) when the observation time $t$ is in the
RD regime respectively in the correlated regime. In addition, a simple exponential decay is encountered
when $t$ is in the saturation regime. If $\mu \gg \nu$, the situation is more complicated, as we do no longer have
the above hierarchy of the length scales, and in many cases no simple dynamical
scaling is observed (see also Fig. 5 in \cite{Chou09b}). Of particular interest
is the case $l_3 > 1/q_{min}$ and $1/q_{max} < l_t < 1/q_{min}$ for which the observation time is in the
correlated regime, as this yields for the global response the expression
\begin{eqnarray}
\chi(t,s)&=&\frac{D}{4\epsilon}\sum_{\mathbf{q}\neq 0}e^{-2q^m\nu(t-s)}\frac{1}{\nu}\left(-\nu s+2q^{m} (\nu s)^2+\cdots\right)\nonumber\\
&=&\frac{DsL^d}{2^{d} \epsilon \pi^{d/2}\Gamma(d/2)}\left\{-\frac{\Gamma(d/m)}{m}\left[2\nu(t-s)\right]^{-d/m} +\cdots\right\}~.
\label{eqchi7}
\end{eqnarray}
This is a remarkable result, as we recover for an arbitrary large perturbation a power-law relaxation and simple dynamical scaling. 
In addition, the exponents describing the scaling behaviour have values that differ from those obtained for a small
perturbation. Indeed, from the expression (\ref{eqchi7}) we obtain $a=\frac{d}{m}-1$ and $\lambda_R = d$. 

For the second protocol, we proceed analogously. We again start from an
initially flat surface and let the system evolve with the strength $D'$ of the noise correlation until the waiting time $s$
at which we set this strength to the final value $D$.
We then compute the function
\begin{equation}
\chi_D(t,s)=\frac{\langle \widetilde{W}^2 \rangle_{D'\rightarrow D}(t,s)- \langle \widetilde{W}^2 \rangle_{D}(t)}{\epsilon}
\label{eqchiD}
\end{equation}
with $\epsilon = D' - D$. This is again a time integrated response. A straightforward calculation yields the expression
\begin{equation}
\chi_D(t,s)= \frac{1}{4 \nu} \sum_{\mathbf{q}\neq 0}\frac{1}{q^m} e^{-2 q^m \nu t} \left( e^{2q^m \nu s} -1 \right)
\label{eqchiD2}
\end{equation}
for this global response. Again, we briefly discuss the limiting cases in the following (see Table \ref{table4}).

1. At early times with $l_t < 1/q_{max}$ we have again a linear decrease of the global response:
\begin{equation}
\chi_D(t,s) \approx \frac{s \pi^{d/2} L^d}{2^{d} \Gamma(\frac{d}{2}) d} \left( 1 - \frac{2d}{d+m} \pi^m \nu t \right)~.
\end{equation}

2. In the correlated regime we obtain for a perturbation ending before the end of the RD regime:
\begin{equation}
\chi_D(t,s)\approx C_{m,d} s t^{-d/m}
\end{equation} 
with
\begin{equation}
C_{m,d} = \frac{\Gamma(\frac{d}{m}) L^d}{2^{d+d/m} \nu^{d/m} \pi^{d/2} m \Gamma(\frac{d}{2})}~.
\end{equation}
If the perturbation only ends in the correlated regime, the result is
\begin{equation}
\chi_D(t,s)= C_{m,d} s t^{-d/m} \left( 1 + \frac{d}{2m} \frac{s}{t} + \cdots \right)~,
\label{eqchiD3}
\end{equation}
yielding the same asymptotic behaviour. This scaling behaviour is shown in Fig. \ref{fig3}e,f for the EW and MH cases.
Comparing the expressions (\ref{eqchiD3}) and (\ref{eqchi3}) we see that
the scaling properties of the response of the square of the surface width depends on how the
growing interface has been perturbed. More precisely we find that (1) the
scaling exponent $a$ is independent of the protocol, but that (2) the autoresponse exponent $\lambda_R$
is now equal to $d$, i.e. the response of the system decays in the long time limit slower when the
perturbation is due to a change of the noise strength. Interestingly,
the global response to a change in $D$ has the same power-law as the global autocorrelation.

3. Finally, if the perturbation continues until the saturation regime is reached, then
\begin{equation}
\chi_D(t,s) \approx \frac{d  L^m}{2^{m+1}\pi^m \nu} e^{- 2 q_{min}^m \nu (t-s)}~.
\end{equation}
If, on the other hand, we measure in the saturation regime the response to a perturbation
that ended before reaching that final regime, we have that
\begin{equation}
\chi_D(t,s) \approx d \, s \, e^{- 2 q_{min}^m \nu t}~.
\end{equation}

\begin{table}[tbp]
\begin{center}
\begin{tabular}{|c||c|c|c|}
\hline
$s$ \textbackslash ~ $t$ & RD& CR & SR \\ \hline\hline
RD & $a_7 \, s \, (1 - c_3 \, t )$ &  $a_8  s^{1-d/m} \, (t/s)^{-d/m}$ & $a_9 \, s\, e^{-b t}$  \\ \hline
CR & $-$ & $a_8  s^{1-d/m} \, (t/s)^{-d/m}$ &  $a_9 \, s \, e^{-b t}$\\ \hline
SR & $-$ &  $-$ &  $\tilde{a}_9 e^{-b (t-s)}$ \\ \hline 
\end{tabular}
\end{center}
\caption{The same as Table \ref{table3}, but now for the response (\ref{eqchiD}) to a change in the strength
of the noise correlation.
The values of the different coefficients can be read off from the equations
in the main text.
}
\label{table4}
\end{table}

\subsection{Global fluctuation-dissipation ratios}
The fluctuation-dissipation ratio \cite{Cri03} has been introduced as a generalisation of the celebrated fluctuation-dissipation
theorem to non-equilibrium systems. Whereas this ratio has some appealing features, as for example its simplicity 
and the possibility to assign an effective temperature to a non-equilibrium system \cite{Cug97}, many studies in model systems
have revealed major problems with that quantity, ranging from observable dependent 
effective temperatures \cite{Fie02,Cal04} to the appearance
of negative temperatures \cite{May06,Gar09}. As a result, the usefulness of the fluctuation-dissipation 
ratio is in general very restricted,
even so it might occasionally yield interesting insights in specific systems.

The fluctuation-dissipation ratios of local quantities and the related effective temperatures have also been studied for
the one-dimensional EW equation \cite{Bus07a} (higher dimensional cases and the MH cases can be inferred from the 
equations given in \cite{Rot06}). This study shows that the ratio of the response $R_\ell(t,s)$ of the height to the
height-height correlation $C_\ell(t,s)$ allows in the linear systems
to characterise the different growth regimes through an effective temperature.

\begin{figure}[h]
\centerline{\epsfxsize=5.25in\ \epsfbox{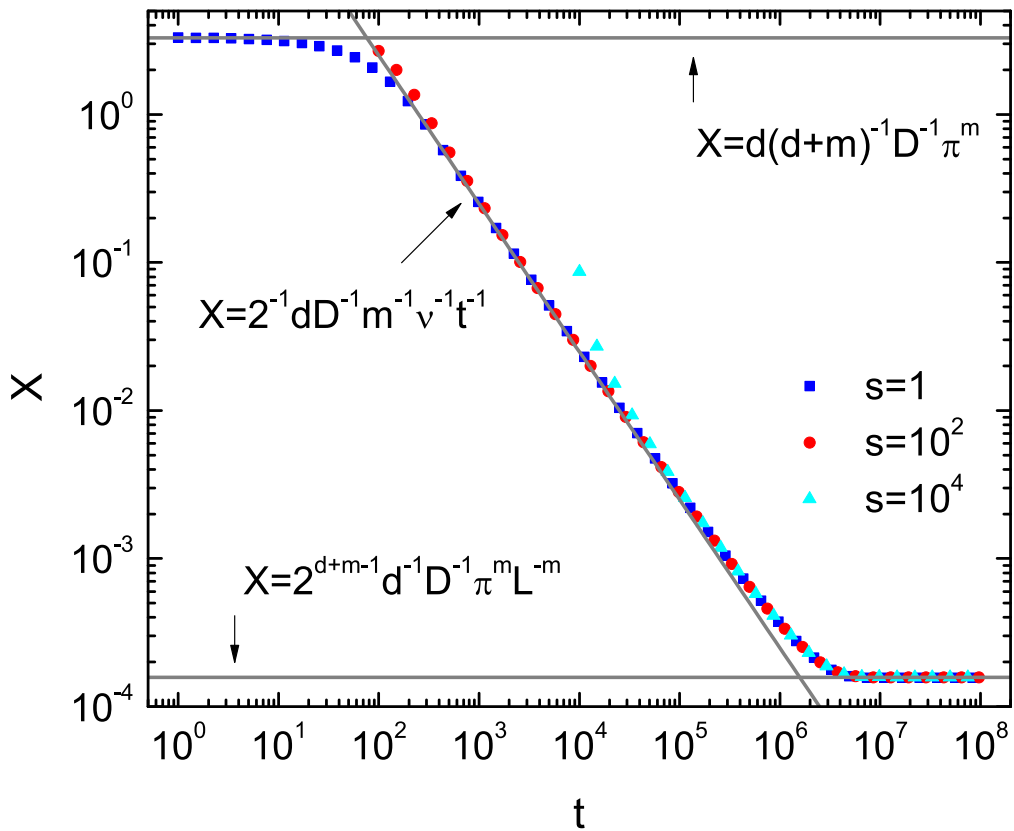}}
\caption{The global fluctuation-dissipation ratio for the one-dimensional EW case, with $m=1$. The ratio (\ref{eq:X}) displays a different 
behaviour in the different regimes. The data have been obtained by numerically evaluating the exact expressions derived in the previous
subsections. The parameters are $L=500$, $D=1$, and $\nu =0.001$.
}
\label{fig4}
\end{figure}

For a time-integrated linear response $\chi(t,s)$ and the conjugate correlation function it is convenient to discuss
the ratio
\begin{equation}
X(t,s) = \frac{\chi(t,s)}{C(t,s)}
\label{eq:X}
\end{equation}
and measure the limit value $X_\infty = \lim_{s\longrightarrow \infty}\lim_{t\longrightarrow \infty} X(t,s)$ \cite{HenPle}.
Strictly speaking, taking this limit does not yield non-trivial results for our global quantities. Assuming this system to be infinite
(such that one remains in the correlated regime in the long-time limit), one immediately sees from the 
Tables \ref{table2} and \ref{table3} that the trivial value $X_\infty = 0$ is obtained for the global fluctuation-dissipation
ratio, and it is not possible to introduce a meaningful effective temperature. 
Still, the quantity $X(t,s)$ does yield at finite times a behaviour that is characteristic for the different regimes, as
shown in Fig. \ref{fig4} for the ratio formed by the global response to a change in $\nu$ and the global correlation. In order to understand
this figure, let us look at a case (the case $s=1$ in the figure) where the waiting time is in the RD regime. If $t \approx s$, i.e. $t$ is also in 
that regime, the ratio (\ref{eq:X}) is constant: $X = \frac{d \pi^m}{D (d+m)}$. 
If $t$ is such that $1/q_{max} < l_t < 1/q_{min}$, i.e. $t$ is in the correlated regime,
$X$ is inversely proportional to $t$: $X = \frac{d}{2 D \nu m} t^{-1}$. The same behaviour, albeit with a different pre-factor, is obtained if we consider
the global response to a change in $D$ instead. Finally, in the saturation regime, with $l_t > 1/q_{min}$, $X$ is again constant, with
$X = \frac{2^{d+m-1} \pi^m}{d D L^m}$. These three regimes are separated by crossover regimes when $l_t \approx 1/q_{max}$ or $l_t \approx 1/q_{min}$.
If we choose a different waiting time, $s= 100$ or $s= 10000$ in Fig. \ref{fig4}, the ratio $X(t,s)$, after some non-universal
behaviour for $t-s \ll s$, rapidly evolves towards the same master curve as the $s=1$ data.

\section{Non-linear growth}
In order to assess the importance of non-linearities, we also studied global quantities derived from the one-dimensional KPZ equation.
We thereby complement the studies \cite{Bus07b,Noh10} which focused on local quantities.

The KPZ equation has been shown to describe faithfully kinetic roughening and to be a paradigmatic model for the description
of a large range of non-equilibrium systems \cite{Hal95}.
This well-known equation is given by the expression
\begin{equation}
\frac{\partial h(\textbf{x},t)}{\partial t}=\nu \nabla^2 h(\textbf{x},t)+\lambda (\nabla h(\textbf{x},t))^2+\eta(\textbf{x},t)
\label{eqKPZ}
\end{equation}
and differs from the EW equation by the non-linear term proportional to the parameter $\lambda$.
Here  $\eta ({\bf x},t)$ is the usual Gaussian white noise with zero mean and
covariance $\left\langle \eta ({\bf x},t)\eta ({\bf y},s)\right\rangle
=D\delta ^d({\bf x}-{\bf y})\delta (t-s)$. 
The numerical integration of (\ref{eqKPZ}) has been the subject of many studies and 
different discretisation schemes have been proposed in order to handle the
non-linearity (see, for example, \cite{New96,Lam98,Buc05,Mir08,Wio10}).

We also studied for this system the global correlation and response functions introduced
in the previous sections. In order to make sure that our results are independent of the integration schemes, we computed
these quantities using different approaches, namely the scheme of Lam and Shin \cite{Lam98} as well as the strong coupling
scheme proposed by Newman and co-workers \cite{New96,New97}. We carefully checked that we are in the correlated regime
at both times $s$ and $t$. We also verified that both approaches yield the same exponents
and scaling functions for the studied quantities \footnote{Due to the very nature of the Newman algorithm, the surface tension is not a parameter that
can be changed in that scheme. Therefore, we only computed with that method the global correlation function as well as the global response to
a change in the noise strength.}. The data obtained from both schemes differ by a non-universal pre-factor,
as expected. 
Small deviations between the data sets are observed for $t \approx s$, but this is again expected as for the Lam/Shin scheme 
we use finite values of $\lambda$, whereas the Newman scheme works in the strong-coupling limit where $\lambda \longrightarrow \infty$.

\begin{figure}[h]
\centerline{\epsfxsize=5.25in\ \epsfbox{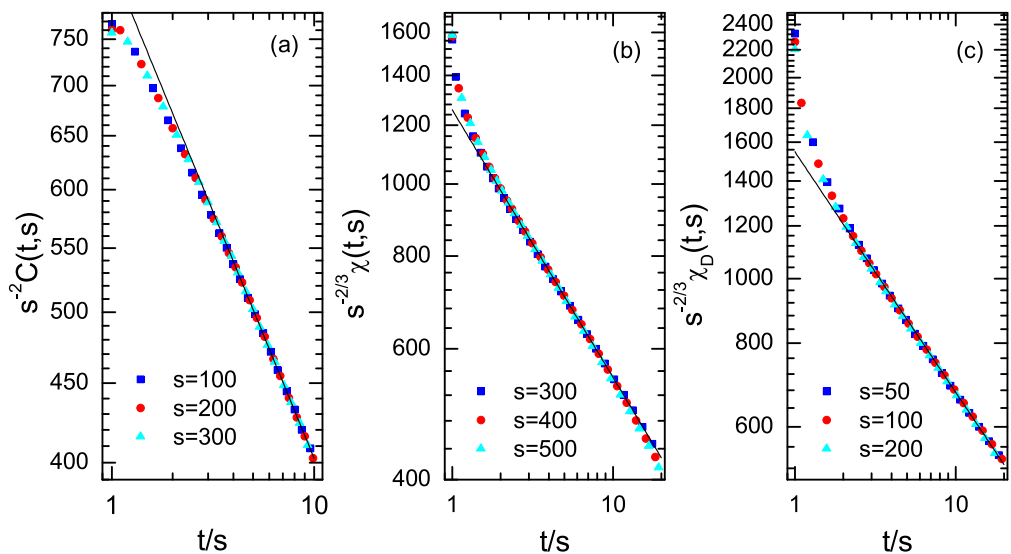}}
\caption{Log-log plots of global quantities as a function of $t/s$ obtained from the one-dimensional KPZ equation in the correlated regime: (a) 
global correlation (b), global response to a change of the surface tension $\nu$, and (c)
global response to a change of the noise strength $D$. The data in (a) and (c) have been obtained with the Newman algorithm \cite{New96,New97},
whereas the data in (b) result from the algorithm of Lam and Shin. We checked that both algorithms produce the same exponents
and scaling functions for a given quantity.
The system size is $L=10000$, with $\lambda =1$, $\nu =1$, and $D=1$. For the response shown in (b),
$\nu$ has the value of 1.1 until time $s$ at which point it is changed to the value 1, whereas for (c) $D$ was changed from the value 1.1 to 1 at time $s$.
The data shown have been obtained after averaging over $5 \times 10^6$ realisations for the correlation and $5 \times 10^5$ realisations
for the response. The lines indicate the asymptotic power-law behaviour.
}
\label{fig5}
\end{figure}

Fig. \ref{fig5} summarizes our main findings regarding the scaling behaviour of global quantities in the one-dimensional
KPZ universality class. For the autocorrelation (\ref{eq:corr}) we find the scaling form
\begin{equation}
C(t,s) = s^2 \, f_C(t/s)
\label{eq:CKPZ}
\end{equation}
with $f_c(y) \sim y^{-1/3}$ for $y \gg 1$. For the response functions we note, and this is in strong contrast to the results we obtained
for the linear equations, that the scaling behaviour does not depend on the protocol that we use for perturbing the system \footnote{
Noting that by rescaling the height $h$ and the time $t$ as $h' = \lambda^{1/3} h$ and $t' = \lambda^{2/3} t$ we can rewrite the
KPZ equation as
\begin{displaymath}
\frac{\partial h'(\textbf{x},t')}{\partial t'}=\nu' \nabla^2 h'(\textbf{x},t')+ (\nabla h'(\textbf{x},t'))^2+\eta(\textbf{x},t')~,
\end{displaymath}
where $\nu' = \nu/(D \lambda^2)^{1/3}$ and the noise is still Gaussian with the same strength $D$, we only consider the same
two types of perturbations as for the linear systems.}. Whether
we change the surface tension or the noise strength, we always end up with the following scaling behaviour of the global response:
\begin{equation}
\chi(t,s) = s^{2/3} \, f_\chi(t/s)
\label{eq:CChi}
\end{equation}
with $f_\chi(y) \sim y^{-1/3}$ for $y \gg 1$. This difference between the global response in the linear and the non-linear cases
is remarkable. At this stage, we can only speculate on the origin of this, but it seems that the restoring mechanism, that is responsible
for the relaxation of the perturbed but still growing interface, originates in the non-linearity, thereby yielding the
same scaling behaviour independent of the nature of the perturbation. Further studies seem to be in order to fully clarify this 
point. 

We can compare the scaling forms (\ref{eq:CKPZ}) and (\ref{eq:CChi}) with themselves as well as with the scaling forms of the local
quantities. We remark that we have: $a = -2/3$ and $a_\ell = -1/3$, $b = -2$ and $b_\ell = -2/3$, $\lambda_C = \lambda_\chi = 1/3$
and $\lambda_{C_\ell} = \lambda_{\chi_\ell} = 2/3$, i.e. the global exponents always differ from their corresponding local exponents.

Finally, let us mention that also for the KPZ case one does not obtain non-trivial limits for global fluctuation-dissipation ratios.
Still, one can perform a similar analysis as done in the previous section for the linear systems, thereby recovering the same three regimes, but with
a different dependence on $t$, $X \sim t^{-4/3}$, in the correlated regime.

\section{Discussion and conclusion}
Our study of global quantities in growth processes has revealed some interesting results, especially in the
correlated regime, and raises a range of open questions that warrant a more in-depth study in the future.

A prediction of the scaling
behaviour of our global quantities is far from obvious, due to the complicated nature of the mean square width
that involves all Fourier components (with the exception of the zero mode) of the Fourier series representation 
of the surface height. We find the available results to be compatible with the scaling form
\begin{equation}
C(t,s) = s^{4 \alpha/z + d/z} \, f_C(t/s)~,
\label{eq:Cscaling}
\end{equation}
with $f_C(y) \approx y^{2 \alpha/z -1}$ for $y \gg 1$, as is readily checked by 
recalling that for the linear equations $\alpha = \frac{z - d}{2}$ and $z = m$, whereas for the
one-dimensional KPZ equation $\alpha = \frac{1}{2}$ and $z = \frac{3}{2}$.
Similarly, for the the global response of the square
of the surface width to a change in the noise strength (we need to specify this 
for the linear systems) is compatible with the scaling form
\begin{equation}
\chi(t,s) = s^{2 \alpha/z} \, f_\chi(t/s)~,
\end{equation}
with $f_\chi(y) \approx y^{2 \alpha/z -1}$ for $y \gg 1$. We are not able to derive these scaling forms at this stage.

In fact, the situation for the response function is rather complex. The reason is of course that for the linear systems
the global response depends on how the system was perturbed, as the different protocols lead to different
scaling forms. This is different for the non-linear KPZ equation, as here the same scaling forms are found for the different
perturbations. Obviously, the non-linearity provides an efficient smoothening mechanism that allows the system to
relax to the state of the unperturbed system in a way that is independent on how the system has been perturbed. This mechanism is
absent in the linear systems, yielding the observed dependence of the response on the protocol. 
A more quantitative study of this effect is left for the future.

Our study also reveals that in the correlated regime the usual limit value of the fluctuation-dissipation ratio involving global quantities 
only leads to trivial results. However, the ratio (\ref{eq:X}) at finite observation times still allows us to distinguish between
the different growth regimes.

In order to measure a global response one needs to perturb in a uniform way the system. On the level of the Langevin equations,
this type of perturbation can only be achieved by varying one of the parameters entering these equations. As we argued in Section 2.3,
based on the microscopic model, the surface tension $\nu$ is directly related to the quantity conjugated to the square of the surface width.
This is different for the strength $D$ of the noise correlator. It is therefore interesting to note the similarities in the response of
the squared width due to a change of either the surface tension or the noise strength. In both cases the non-equilibrium exponents that
govern the behaviour in the ageing regime are not identical to the corresponding exponents describing the correlation function of the squared
width. This seems therefore to be a general property of global two-time quantities.

Experimentally a global response is much easier to measure than a local one. Local perturbations and local measurements
are not easily achieved in a growing system. This is different for a global response that is readily observed when
changing the experimental conditions, for example through a change in temperature, as this automatically affects the
whole system. Another advantage of studying global quantities in the context of growth processes is seen at the upper
critical dimension. Indeed, global quantities continue to display a standard dynamical scaling behaviour at the
critical dimension. Local quantities, however, generically display a logarithmic dependence on time in that case which
makes an analysis much more demanding. For these reasons we hope that our study will motivate other groups, especially
experimental ones, to use global quantities for the characterisation of non-equilibrium growth processes.

\ack
We thank Sebastian Bustingorry for bringing Reference \cite{Ngu09} to our attention
and Uwe T\"{a}uber for an enlightening discussion on conjugated
quantities within a field-theoretical setting.
This work was supported by the US National
Science Foundation through grant DMR-0904999.


\section{References}
\begin{thebibliography}{999}
\bibitem{Mea93} Meakin P 1993 {\it Phys.\ Rep.} {\bf 235} 189
\bibitem{Hal95} Halpin-Healy T and Zhang Y-C 1995 {\it Phys.\ Rep.} {\bf 254} 215
\bibitem{Bar95} Bar\'{a}basi A-L and Stanley H E 1995 {\it Fractal Concepts in Surface Growth}
(Cambridge: Cambridge University Press).
\bibitem{Kru97} Krug J 1997  {\it Adv. Phys.} {\bf 46} 139 
\bibitem{Kru95} Krug J 1995 in {\it Scale Invariance, Interfaces, and Non-Equilibrium
Dynamics}, edited by A. McKane {\it et al.} (New York: Plenum, New York)
\bibitem{1985FF} Family  F and Vicsek T 1985 {\it J. Phys. A} {\bf 18} L75
\bibitem{Chou09a} Chou Y-L and Pleimling M 2009 {\it Phys. Rev. E} {\bf 79} 051605
\bibitem{Kal99} Kallabis H and Krug J 1999 {\it Europhys. Lett.} {\bf 45} 20
\bibitem{Rot06} R\"{o}thlein A, Baumann F and Pleimling M 2006 {\it Phys. Rev. E} {\bf 74} 061604
\bibitem{Rot07} R\"{o}thlein A, Baumann F and Pleimling M 2007 {\it Phys. Rev. E} {\bf 76} 019901(E)
\bibitem{Bus07a} Bustingorry S, Cugliandolo L F and Iguain J L 2007 {\it J. Stat. Mech.} P09008
\bibitem{Bus07b} Bustingorry S 2007 {\it J. Stat. Mech.} P10002
\bibitem{Chou09b} Chou Y-L, Pleimling, M, and Zia R K P 2009 {\it Phys. Rev. E} {\bf 80} 061602
\bibitem{Ngu09} Nguyen T T T, Bonamy D, Phan Van L, Cousty J, and Barbier L 2009 arXiv:0904.2200
\bibitem{Noh10} Henkel M, Noh J D, and Pleimling M 2010 unpublished
\bibitem{HenPle} Henkel M and Pleimling M 2010 {\it Non-equilibrium phase transitions 
Volume 2: Ageing and dynamical scaling far from equilibrium} (Heidelberg: Springer)
\bibitem{May03} Mayer P, Berthier L, Garrahan J P, and Sollich P 2003 {\it Phys. Rev. E} {\bf 68} 016116
\bibitem{Ple05} Pleimling M and Gambassi A 2005 {\it Phys. Rev. B} {\bf 71} 180401(R)
\bibitem{Ann06} Annibale A and Sollich P 2006 {\it J. Phys A: Math. Gen.} {\bf 39} 2853
\bibitem{Cal06} Calabrese P, Gambassi A, and Krzakala F 2006 {\it J. Stat. Mech.} P06016
\bibitem{Dut09} Dutta S B, Henkel M, and Park H 2009 {\it J. Stat. Mech.} P03023
\bibitem{Edw82} Edwards S F and Wilkinson D R 1982 {\it Proc. R. Soc. London Ser. A}
{\bf 381} 17
\bibitem{Mul63} Mullins W W 1963 in {\it Metal surfaces: Structure, energetics, and kinetics}
(Metals Park, Ohio: Am. Soc. Metal)
\bibitem{Kar86} Kardar M, Parisi G, and Zhang Y-C 1986 {\it Phys. Rev. Lett.} {\bf 56} 889
\bibitem{1986FF} Family F 1986 {\it J. Phys. A} {\bf 19} L441
\bibitem{Sar96} das Sarma S, Lanczycki C J, Kotlyar R, and 
Ghaisas S C 1986 {\it Phys. Rev. E} {\bf 53} 359
\bibitem{Men96}  Meng B and Weinberg W H 1996 {\it Surface Science} {\bf 364} 151
\bibitem{Bar99}  Bartelt M C and Evans J W 1999 {\it Surface Science} {\bf 423} 189
\bibitem{Liu05}  Liu Z-J and Shen Y G 2005 {\it Surface Science} {\bf 595} 20
\bibitem{Els04}  Elsholz F, Sch\"{o}ll E and Rosenfeld A 2004 {\it Appl. Phys.
Lett.} {\bf 84} 4167
\bibitem{Gha01}  Ghaisas S V 2001 {\it Phys. Rev. E} {\bf 63} 062601
\bibitem{Bar96} Barbier L, Masson L, Cousty J, and Salanon B 1996 {\it Surface Science} 
{\bf 345} 197
\bibitem{Dou02} Dougherty D B, Lyubinetsky I, Williams E D, Constantin M, Dasgupta C, and Das Sarma S
2002 {\it Phys. Rev. Lett.} {\bf 89} 136102 
\bibitem{Dou05} Dougherty D B, Tao C, Bondarchuk O, Cullen W G, Williams E D,
Constantin M, Dasgupta C, and Das Sarma S 2005 {\it Phys. Rev. E} {\bf 71} 021602
\bibitem{Bon05} Bondarchuk O, Dougherty D B, Degawa M, Williams E D,
Constantin M, Dasgupta C, and Das Sarma S 2005 {\it Phys. Rev. B} {\bf 71} 045426
\bibitem{Hor01}  Horowitz C M, Monetti R A and Albano E V {\it Phys. Rev.
E} {\bf 63} 066132
\bibitem{Cha02}  Chame A and Aar$\tilde{a}$o Reis F D A 2002 {\it Phys. Rev. E}
{\bf 66} 051104
\bibitem{Hor03}  Horowitz C M and Albano E V 2003 {\it Eur. Phys. J. B} {\bf 31} 563
\bibitem{Kol04}  Kolakowska A, Novotny M A and Verma P S 2004 {\it Phys. Rev. E}
{\bf 70} 051602
\bibitem{Mur04}  Muraca D, Braunstein L A and Buceta B C 2004 {\it Phys. Rev. E}
{\bf 69} 065103(R)
\bibitem{Bra05}  Braunstein L A and Lam C-H 2005 {\it Phys. Rev. E} {\bf 72} 026128
\bibitem{Iru05}  Irurzun I, Horowitz C M and Albano E V 2005 {\it Phys. Rev. E}
{\bf 72}, 036116
\bibitem{Hor06}  Horowitz C M and Albano E V 2006 {\it Phys. Rev. E} {\bf 73} 031111
\bibitem{Rei06}  Aar$\tilde{a}$o Reis F D A 2006 {\it Phys. Rev. E} {\bf 73} 021605
\bibitem{Oli06}  Oliveira T J, Dechoum K, Redinz J A and Aar$\tilde{a}$o Reis F D A 
2006 {\it Phys. Rev. E} {\bf 74} 011604
\bibitem{Kol06}  Kolakowska A, Novotny M A and Verma P S 2006 {\it Phys. Rev. E}
{\bf 73}, 011603
\bibitem{Maj96} Majaniemi S, Ala-Nissila T, and Krug J 1996 {\it Phys. Rev. B} {\bf 53} 8071
\bibitem{Gie01} Giesen M 2001 {\it Prog. Surf. Sci.} {\bf 68} 1
\bibitem{Dou04} Dougherty D B, Lyubinetsky I, Einstein T L and Williams E D 2004
{\it Phys.\ Rev.\ B} {\bf 70} 235422
\bibitem{Wol90} Wolf D E and Villain J 1990 {\it Europhys.\ Lett.} {\bf 13} 389 
\bibitem{Gol91} Golubov\'{i}c L and Bruinsma R 1991 {\it Phys. Rev. Lett.} {\bf 66} 321
\bibitem{Das91} Das Sarma S and Tamborenea P 1991 {\it Phys. Rev. Lett.} {\bf 66} 325
\bibitem{Igu09} Iguain J L, Bustingorry S, Kolton A B and Cugliandolo L F 2009
{\it Phys. Rev. B} {\bf 80} 094201
\bibitem{Cri03} Crisanti A and Ritort F 2003 {\it J. Phys A: Math. Gen.} {\bf 36} R181
\bibitem{Cug97} Cugliandolo L F, Kurchan J, and Peliti L 1997 {\it Phys. Rev. E} {\bf 55} 3898
\bibitem{Fie02} Fielding S and Sollich P 2002 {\it Phys. Rev. Lett.}  {\bf 88} 050603
\bibitem{Cal04} Calabrese P and Gambassi A 2004 {\it J. Stat. Mech.} P07013
\bibitem{May06} Mayer P, Berthier L, Garrahan J P and Sollich P 2006 {\it Phys. Rev. Lett.}  {\bf 96} 030602
\bibitem{Gar09} Garriga A, Pagonabarraga I, and Ritort F 2009 {\it Phys. Rev. E} {\bf 79} 041122
\bibitem{New96} Newman T J and Bray A J 1996 {\it J. Phys. A: Math. Gen} {\bf 29} 7917
\bibitem{Lam98} Lam C-H and Shin F G 1998 {\it Phys. Rev. E} {\bf 58} 5592
\bibitem{Buc05} Buceta R C 2005 {\it Phys. Rev. E} {\bf 72} 017701
\bibitem{Mir08} Miranda V G and Aarão Reis F D A 2008 {\it Phys. Rev. E} {\bf 77} 031134
\bibitem{Wio10} Wio H S, Revelli J A, Deza R R, Escudero C and de la Lama M S 2010 {\it EPL} {\bf 89} 40008
\bibitem{New97} Newman T J and Swift M R 1997 {\it Phys. Rev. Lett.} {\bf 79} 2261
\end{thebibliography}
\end{document}